\documentclass[twocolumn,showpacs,aps,pre]{revtex4}
\usepackage{makeidx}
\usepackage{amsmath}
\usepackage{amssymb}
\usepackage{graphicx}
\usepackage{graphics}
\begin{document}
\title{Statistical mechanical description of liquid systems in electric field}
\author{Semjon Stepanow and Thomas Thurn-Albrecht}
\affiliation{Institut f\"{u}r Physik, Martin-Luther-Universit\"{a}t Halle-Wittenberg,
D-06099 Halle, Germany}
\date{\today}
\begin{abstract}
We formulate the statistical mechanical description of liquid systems for both polarizable and polar systems in an electric field in the $\mathbf{E}$-ensemble, which is the pendant to the thermodynamic description in terms of the free energy at constant potential. The contribution of the electric field to the configurational integral $\tilde{Q}_{N}(\mathbf{E})$ in the $\mathbf{E}$-ensemble is given in an exact form as a factor in the integrand of $\tilde{Q}_{N}(\mathbf{E})$. We calculate the contribution of the electric field to the Ornstein-Zernike formula for the scattering function in the $\mathbf{E}$-ensemble. As an application we determine the field induced shift of the critical temperature for polarizable and polar liquids, and show that the shift is upward for polarizable liquids and downward for polar liquids.
\end{abstract}

\maketitle


\section{Introduction}\label{intro}
The behavior of liquid systems in an external electric field has been the
subject of experimental and theoretical interest over many decades \cite{kirkwood36}-\cite{hegseth-prl04}.
A recent example are block copolymer melts which are of particular interest due to field induced uniform macroscopic alignment in the microphase separated state \cite{amundson-ma93}-\cite{matsen06} (and references therein), which is of basic importance for applications using self assembled block copolymer structures for patterning and templating of nanostructures \cite{park03}. A more basic question concerns the shift of the critical temperature for low molecular weight systems in  an electric field. Although this issue has been the subject of investigations over many years \cite{debye-kleboth65}-\cite{hegseth-prl04}, a microscopic explanation of the shift has not been provided.
 Most of the existing theoretical work on the contribution of the electric field to thermodynamic quantities is based on a phenomenological description in the framework of macroscopic electrodynamics.
It is well-known \cite{landau-lifshitz8} that for thermodynamics in an electric field one should distinguish between thermodynamic potentials at constant dielectric displacement or constant charges $F(T,V,\mathbf{D})$ and at constant electric field or potential $\tilde{F}(T,V,\mathbf{E})$. However, the available theoretical studies do not provide the tools for calculating the thermodynamic quantities and the correlation functions (in particular the scattering function) in the ensemble at constant electric field, which is of primary experimental relevance.
In this Letter we address this problem and establish the statistical mechanical description of liquid systems in electric fields at constant electric field $\tilde{F}(T,V,\mathbf{E})$ by introducing the $T-V-\mathbf{E}$-ensemble following the analogy between the $T-V$ and $T-p$ ensembles of Statistical Mechanics \cite{kubo-book}. The contribution of the electric field to the configurational integral $\tilde{Q}_{N}(\mathbf{E})$ in the $T-V-\mathbf{E}$-ensemble is given in an exact form as a factor in the integrand of $\tilde{Q}_{N}(\mathbf{E})$. This $\mathbf{E}$-ensemble
provides a systematic way for the calculation of thermodynamic quantities, correlation functions in electric fields, and among others the calculation of the dielectric constant. We also derive the contribution of the electric field to the Ornstein-Zernike expression of the scattering function to quadratic order in atomic polarizabilities in both $\mathbf{E}_{0}$- and $\mathbf{E}$-ensemble. Using these results we consider the shift of the critical temperature for both polarizable and polar liquids in an electric field, and give an explanation for the different sign of the shift in these systems.

The article is organized as follows. Section \ref{polar-dip} introduces to the formalism of the description of simple liquids in an electric field. Section \ref{perm-dip} generalizes the formalism to polar systems. Section \ref{T_c-shift} considers the shift of the critical temperature in simple and polar liquids, and the comparison with experimental results.

\section{Statistical Mechanics of simple liquids in an electric field}\label{theory}

\subsection{Description of polarizable systems in the $\mathbf{E}$-ensemble} \label{polar-dip}

Let us first review the statistical mechanics of a system of interacting molecules without permanent electric dipole moments with Hamiltonian $H_0$ in an external electric field $\mathbf{E}_{0}$. The induced dipole moment of the $ith$ molecule for a system of molecules in
an external electric field $E_{0}$ is given by \cite{isihara62}
\begin{equation}
p_{i}=\sum\limits_{j}\left[ 1+\alpha _{i}T\right] _{ij}^{-1}\alpha
_{j}E_{0}(r_{j}),  \label{e8}
\end{equation}%
where the summation occurs over all molecules, and the molecular
polarization tensor $\alpha $ is defined by $p_{i}^{\alpha }=\alpha
_{i}^{\alpha \beta }E_{0}^{\beta }(\mathbf{r}_{i})$ with $a^{\alpha \beta }=\alpha _{s}\delta ^{\alpha \beta }$. The tensor of
dipole-dipole interactions $T_{ji}^{\alpha \beta }$ is given by
\begin{equation*}
T^{\alpha \beta }(\mathbf{r}_{j}-\mathbf{r}_{i})=-\nabla _{j}^{\alpha
}\nabla _{j}^{\beta }\frac{1}{r_{ji}}=\frac{\delta ^{\alpha \beta }}{%
r_{ji}^{3}}-3\frac{r_{ji}^{\alpha }r_{ji}^{\beta }}{r_{ji}^{5}}.
\end{equation*}%
It is assumed in (\ref{e8}) and in the following that the diagonal elements $%
T_{ii}$ are zero. The total microscopic strength of the electric field at
the position $\mathbf{r}$ is given by the sum of the external field $\mathbf{%
E}_{0}(\mathbf{r})$ and the field of the induced dipoles $\mathbf{p}_{j}$%
\begin{equation}
E^{\alpha }(\mathbf{r})=E_{0}^{\alpha }(\mathbf{r})-\sum\limits_{j}T^{\alpha
\beta }(\mathbf{r}-\mathbf{r}_{j})p_{j}^{\beta }.  \label{E-m}
\end{equation}%
The total interaction energy of the induced dipoles is given by \cite{isihara62}
\begin{eqnarray}
H_{pol} &=&-\frac{1}{2}\sum\limits_{i,j}E_{0}(r_{i})\left[ 1+\alpha _{i}T%
\right] _{ij}^{-1}\alpha _{j}E_{0}(r_{j})  \notag \\
&=&-\frac{1}{2}\sum\limits_{i}p_{i}^{\alpha }E_{0}^{\alpha }(r_{i}).
\label{e12}
\end{eqnarray}%
The last line in (\ref{e12}) corresponds to the electric energy of the
induced dipole moments $p_{i}^{\alpha }$ given by Eq.~(\ref{e8}) in the
external electric field $E_{0}^{\alpha }(r_{i})$. The expansion of $H_{pol}$
in Eq.~(\ref{e12}) to linear order in \thinspace $T$ yields%
\begin{equation}
H_{pol}\simeq -\frac{1}{2}\sum\limits_{i}E_{0}\alpha E_{0}+\frac{1}{2}%
\sum\limits_{i,j}E_{0}\alpha _{i}T(\mathbf{r}_{i}-\mathbf{r}_{j})\alpha
_{j}E_{0}+\cdots .  \label{H_el_exp}
\end{equation}%
The Cartesian indices are suppressed in (\ref{H_el_exp}). The configuration
integral in the electric field is given by%
\begin{equation}
Q_{N}(E_0)=\int d^{3}r_{1}\cdots \int d^{3}r_{N}e^{-\beta H_{0}-\beta H_{pol}},
\label{vw2}
\end{equation}%
where $H_{0}$ is the Hamiltonian which contains the repulsive hard-core and
the attractive van der Waals interactions between the molecules. Eq.~(\ref{vw2})
allows the statistical mechanical computation of the free energy $F$ in a
constant external electric field $\mathbf{E}_{0}$ which can be identified
with the dielectric displacement $\mathbf{D}$, because both obey the same
Maxwell equation $\mathrm{div}\mathbf{E}_{0}=4\pi \rho _{ext}$.

The average of $H_{pol}$ given by Eq.~(\ref{H_el_exp}) over positions of
molecules results in
\begin{eqnarray}
\left\langle H_{pol}\right\rangle  &=&-\frac{N}{2}\alpha _{s}\mathbf{E}%
_{0}^{2}+\frac{N^{2}}{V}\alpha _{s}^{2}4\pi \int d^{3}q\frac{q^{\alpha
}q^{\beta }}{q^{2}}\delta ^{(3)}(\mathbf{q})E_{0}^{\alpha }E_{0}^{\beta }
\notag \\
&\simeq &-\frac{1}{2}\alpha _{s}N\mathbf{E}_{0}^{2}+\frac{N^{2}}{V}\alpha
_{s}^{2}\frac{4\pi }{3}\mathbf{E}_{0}^{2}.  \label{vw7}
\end{eqnarray}%
To calculate the integral over $q$ in the first line of Eq.~(\ref{vw7}) we replaced the delta function $\delta
^{(3)}(\mathbf{q})$ by the normalized bell-shaped function $\delta _{a}(%
\mathbf{q})=(1/2\pi a^{2})^{3/2}\exp (-(1/2a^{2})\mathbf{q}^{2})$. The free energy is related to $Q_{N}$ by $\mathcal{F}=F_{0}-k_{B}T\ln Q_{N}/V^{N}$, and does not contain the energy of the external electric field. The expansion of $Q_{N}$ in (\ref{vw2}) up to linear terms in dipole-dipole interactions and the use of (\ref{vw7}) results in the following expression of the free energy, which is exact to 2nd order in $\alpha _{s}$
\begin{equation}
\mathcal{F}=F_{0}-\frac{N\alpha _{s}}{2}E_{0}^{2}\left( 1-\frac{8\pi }{3}%
\alpha _{s}\frac{N}{V}+\cdots \right) .  \label{vw8}
\end{equation}%

In thermodynamics the free energies  at constant dielectric displacement $F(T,V,\mathbf{D})$ and at constant electric field $\tilde{F}(T,V,\mathbf{E})$ are related by a Legendre transform as $\tilde{F}=F-\mathbf{ED}V/4\pi $ with the
differentials given by \cite{landau-lifshitz8}
\begin{eqnarray}
dF&=&-SdT-pdV+\frac{V}{4\pi }\mathbf{E}d\mathbf{D},  \notag \\
d\tilde{F}&=&-SdT-pdV-\frac{V}{4\pi }\mathbf{D}d\mathbf{E},  \label{td2}
\end{eqnarray}
where for simplicity homogeneous fields are assumed in the above expressions.

We will now consider the question of the statistical mechanical calculation
of the thermodynamic quantities in the ensemble at constant potential. According to
the general principles of construction of different ensembles in Statistical Mechanics \cite{kubo-book} the
relation between $F$ and $\tilde{F}$ should correspond to the transition from the $T-V-\mathbf{E}_{0}$-ensemble to a $T-V-\mathbf{E}
$-ensemble. Following the known examples in Statistical Mechanics \cite%
{kubo-book}, as for example the relation between the $T-V$ and $T-p$ ensembles,
we define the $T-V-\mathbf{E}$ ensemble by the following expression for the
configurational integral
\begin{eqnarray}
\tilde{Q}_{N}(\mathbf{E}) &=&\int D\mathbf{E}_{0}(\mathbf{r})\exp \left(
\beta \int d^{3}r\frac{\mathbf{E}\mathbf{E}_{0}}{4\pi }\right)  \notag \\
&\times &\exp \left( -\beta \int d^{3}r\frac{\mathbf{E}_{0}^{2}}{8\pi }%
\right) Q_{N}(\mathbf{E}_{0}),  \label{td5}
\end{eqnarray}%
where the term $\sim \mathbf{E}_{0}^{2}$ corresponds to the energy of the
external electric field. The integration over the field strength in (\ref%
{td5}) occurs at every $\mathbf{r}$, i.e. (\ref{td5}) is a functional
integral.

Indeed, the functional integral in (\ref{td5}) is Gaussian, and consequently
the integration over $\mathbf{E}_{0}(\mathbf{r})$ can be performed exactly using the quadratic complement. We obtain
\begin{eqnarray}
&&\tilde{Q}_{N}(\mathbf{E})= \int d\Gamma \exp \left( -\beta H_{0}\right)
\exp \left[ \frac{1}{2}\ln (\frac{8\pi ^{2}}{\beta }\det A^{-1})\right]
\notag \\
&\times &\exp \left[ \frac{\beta }{8\pi }\int d^{3}r\int d^{3}r^{\prime
}E(r)A^{-1}(r,r^{\prime })E(r^{\prime })\right] ,  \label{td5c}
\end{eqnarray}%
where $d\Gamma $ denotes here integrations over positions $r_{1},\cdots
,r_{N}$ of the particles, and the matrix $A$ is defined by%
\begin{equation}
A(r,r^{\prime })=\delta (r-r^{\prime })-4\pi n(r)(I+\alpha Tn)_{r,r^{\prime
}}^{-1}\alpha ,  \label{td5d}
\end{equation}%
where $I\rightarrow \delta (r-r^{\prime })$ is the identity matrix, and $n(r)
$ is the microscopic density $n(r)=\sum\limits_{i}\delta (r-r_{i})$. The
Cartesian indices of $A$ are suppressed. The expansion of $%
A^{-1}(r,r^{\prime })$ under the integral in the last line of (\ref{td5c})
to the second order in atomic polarizability gives rise to a factor $\exp (-\beta
W_{2,pol})$ with%
\begin{eqnarray}
W_{2,pol} &=&-\frac{1}{2}\sum\limits_{i,j}E^{\alpha }\alpha _{i}^{\alpha
\beta }\left[ 4\pi \delta ^{\beta \mu }\delta (\mathbf{r}_{i}-\mathbf{r}%
_{j})\right.   \notag \\
&-&\left. T^{\beta \mu }(\mathbf{r}_{i}-\mathbf{r}_{j})\right] \alpha
_{j}^{\mu \gamma }E^{\gamma }.  \label{W2,el}
\end{eqnarray}%
The first term in (\ref{W2,el}) originates from interactions of induced
dipoles with the electric field to the 2nd order, while the second term is
the 1st order contribution of the dipole-dipole interactions. For homogeneous field the term linear in $\alpha _{s}$ does not depend on the positions of the molecules, and consequently does not contribute to
observables. Eq.~(\ref{td5c}) is the starting point for a calculation of the
dielectric constant by summations of subseries of perturbation expansions in
powers of the atomic polarizabilities. It is easy to see that the
preaveraging of $A(r,r^{\prime })$ in the exact expression (\ref{td5d}) according to $A(r,r^{\prime })\rightarrow \langle A(r,r^{\prime })\rangle =
\delta (r-r^{\prime })\left( 1-4\pi n\alpha \left( 1+8\pi \alpha n/3\right)
^{-1}\right) $ results in the following expression for $\tilde{Q}_{N}(\mathbf{E})=\exp(-\beta\tilde{F}) = \exp\left(-\beta F_0 +\beta\int\varepsilon \mathbf{E}^2(\mathbf{r})d^3r/8\pi\right)$ with the dielectric
constant $1/\varepsilon =1-4\pi n\alpha /(1+8\pi n\alpha /3),$ which is
equivalent to the Clausius-Mossotti relation. Effects of the electric field beyond $\mathbf{E}^{2}$ can be also studied using (\ref{td5c}).
The $\mathrm{ln \,det}$ term in (\ref{td5c}) does not depend on the electric field but on the positions of the molecules. It has the following
interpretation: Induced dipoles appear also in the absence of an external
electric field due to thermal fluctuations. Their contribution to the
interaction energy between the molecules can be absorbed into interactions between the molecules which are independent of the external electric field, and are out of interest in the present work.

We now will establish the contribution of the electric field to the static
scattering function in the $\mathbf{E}_0$-ensemble. We start with the Ornstein-Zernike equation \cite{simple-liquids68}
\begin{equation*}
h(\mathbf{r}_{1}-\mathbf{r}_{2})=c(\mathbf{r}_{1}-\mathbf{r}_{2})+n\int
d^{3}r_{3}c(\mathbf{r}_{1}-\mathbf{r}_{3})h(\mathbf{r}_{3}-\mathbf{r}_{2}),
\end{equation*}
which expresses the correlation function $h(\mathbf{r}
_{1}-\mathbf{r}_{2})=g(\mathbf{r}_{1}-\mathbf{r}_{2})-1$ through the direct
correlation function $c(\mathbf{r}_{1}-\mathbf{r}_{2})$,
where $n=N/V$ is the average density, and the distribution function $g(%
\mathbf{r} _{1}-\mathbf{r}_{2})$ is defined by
\begin{equation}
g(\mathbf{r}_{1}-\mathbf{r}_{2})=\frac{V^{2}\int d^{3}r_{3}\cdots \int
d^{3}r_{N}e^{-\beta H_{0}-\beta H_{pol}}}{\int d^{3}r_{1}\cdots \int
d^{3}r_{N}e^{-\beta H_{0}-\beta H_{pol}}}.  \label{vw17}
\end{equation}
In the case of homogeneous field we obtain from (\ref{vw17}) to the lowest order in dipole-dipole interactions
\begin{equation*}
g(\mathbf{r}_{1}-\mathbf{r}_{2}) \approx g_{0}(\mathbf{r}_{1}-\mathbf{r}%
_{2})-\beta E_{0}\alpha T( \mathbf{r}_{1}-\mathbf{r}_{2})\alpha E_{0}.
\label{vw21}
\end{equation*}
The scattering function is expressed through the Fourier transform of the
direct correlation function as \cite{simple-liquids68} $S^{-1}(\mathbf{q}%
)=1-nc(\mathbf{q})$. Using the above relations and the Ornstein-Zernike formula we obtain the scattering function in the electric field $\mathbf{E}_0$ in the vicinity of the critical point by taking into account the interactions of induced dipoles to lowest order as
\begin{equation}
S^{-1}(\mathbf{q})=\tau +cq^{2}+4\pi \beta n\alpha _{s}^{2}\frac{\left(
\mathbf{qE}_{0}\right) ^{2}}{q^{2}}+\cdots ,  \label{vw28}
\end{equation}
where $\tau =T-T_c$.

To establish the contribution of the electric field to the scattering
function in the $\mathbf{E}$-ensemble we start with the expression of the radial distribution function
\begin{eqnarray}
&&\tilde{g}(\mathbf{r}_{1},\mathbf{r}_{2})=\frac{V^{2}}{\tilde{Q}_{N}(
\mathbf{E})}\int D\mathbf{E}_{0}(\mathbf{r})\exp \left( \beta \int d^{3}r
\frac{\mathbf{E}(\mathbf{r})\mathbf{E}_{0}(\mathbf{r})}{4\pi }\right. \notag \\
&&\left. -\beta \int d^{3}r\frac{\mathbf{E}_{0}^{2}(\mathbf{r})}{8\pi }
\right) \int d^{3}r_{3}\cdots \int d^{3}r_{N}e^{-\beta H_{0}-\beta H_{pol}},
\label{vw32}
\end{eqnarray}
which is in accordance with the definition of the partition function in Eq.~(\ref%
{td5}). Carrying out the functional integration over $\mathbf{E}_{0}(\mathbf{%
r})$ similarly to Eq.~(\ref{td5c}) we arrive at
\begin{eqnarray}
&&\tilde{g}(\mathbf{r}_{1},\mathbf{r}_{2})=\frac{1}{\tilde{Q}_{N}(\mathbf{E}%
) }V^{2}\int d^{3}r_{3}\cdots \int d^{3}r_{N}e^{-\beta H_{0}}  \notag \\
&\times &e^{\frac{1}{2}\ln \frac{8\pi ^{2}}{\beta }\det A^{-1}+\frac{\beta }{
8\pi }\int d^{3}r\int d^{3}r^{\prime }E(r)A^{-1}(r,r^{\prime })E(r^{\prime
})},  \label{vw32a}
\end{eqnarray}
where $\tilde{Q}_{N}(\mathbf{E})$ and $A(r,r^{\prime })$ are defined by Eqs.~(\ref{td5c}) and (\ref{td5d}),
respectively. The lowest-order contribution to the direct correlation function from the electric field can be obtained by expanding $A^{-1}(r,r^{\prime})$ in Eq.~(\ref{vw32a}) similar to the corresponding expansion of Eq.~(\ref{td5c}), which results in
\begin{equation}
\tilde{g}(\mathbf{r}_{1},\mathbf{r}_{2})\simeq g_{0}(\mathbf{r}_{1}-\mathbf{%
r }_{2})\exp \left( -\beta W_{2,pol}\right),  \label{vw34}
\end{equation}
where the effective binary energy $W_{2,pol}$ is given by Eq.~(\ref{W2,el}).
Following the same steps as in the above derivation in the $\mathbf{E}_{0}$%
-ensemble we obtain the scattering function in the $\mathbf{E}$-ensemble in homogeneous field as
\begin{equation}
\tilde{S}^{-1}(\mathbf{q})=\tau +cq^{2}-2\pi \beta n\alpha _{s}^{2}\left(
\mathbf{E}^{2}-\frac{\left( \mathbf{qE}\right) ^{2}}{q^{2}}\right) +\cdots .
\label{vw35}
\end{equation}
The isotropic term on the right-hand side of (\ref{vw35}) is due to
interactions of the induced dipoles with the fluctuating electric field to
order $\alpha _{s}^{2}$, while the anisotropic term is associated with the
dipole-dipole interactions. The comparison of the electric-field
contribution with the term $\tau =T-T_{c}$ in (\ref{vw35}) yields that the
critical temperature in an electric field is shifted upwards, which is in
accordance with the thermodynamic consideration for the van der Waals gas in the next paragraph. The density fluctuations become anisotropic in an electric field, such that the fluctuations with wave vectors $\mathbf{q}_{\bot }$ transverse to the field strength will be enforced, while fluctuations with the wave vectors $\mathbf{q}_{\Vert }$ parallel to $\mathbf{E}$ remain unchanged. The upward shift of $T_c$ is due to the isotropic term, which is absent in the $\mathbf{E}_0$-ensemble. The anisotropic character of the density fluctuations described by Eq.~(\ref{vw35}) has the consequence that the instability limit for fluctuations with wave vectors transverse to $%
\mathbf{E}$ will be reached at higher temperatures, which also means that an ordered state with the interface parallel to the electric field will be
preferred. Note that the anisotropic term in (\ref{vw35}) has the same form as the uniaxial anisotropy in magnetic systems \cite{aharony76}, and can
change the upper critical dimension from $d_c=4$ to $d_c=3$.

As a last point in this subsection we will check the validity of the relation between the compressibility and the structure factor
\begin{equation}
1+n\int \tilde{h}(r)d^{3}r=\tilde{S}(q=0)=k_{B}T\left( \frac{\partial n}{
\partial \tilde{p}}\right) _{T}  \label{vw29}
\end{equation}
in an electric field in the $\mathbf{E}$-ensemble at the critical point. The computation of $\left(\partial \tilde{p}/\partial n\right) _{T}$ using Eq.~(\ref{vw15a}) in the next Section yields at
the critical point $ \left(\partial \tilde{p}/\partial n\right) _{T_c}=-\frac{4\pi }{3}
\alpha _{s}^{2}n_{c}\mathbf{E}^{2}$.
The latter should be compared with $\tilde{S}^{-1}(q\rightarrow
0)_{\left\vert T=T_{c}\right. }$ which is obtained from Eq.~(\ref{vw35}) as
\begin{equation*}
\tilde{S}^{-1}(q\rightarrow 0)_{\left\vert T=T_{c}\right. }=-2\pi \beta
_{c}n_{c}\alpha _{s}^{2}\mathbf{E}^{2}+2\pi \beta _{c}n_{c}\alpha
_{s}^{2}t^{\mu }t^{\nu }E^{\mu }E^{\nu } .  \label{vw31}
\end{equation*}
The average of the right-hand side over the directions of the unit vector $t^{\mu }=q^{\mu }/q$ gives the factor $\delta^{\mu \nu}/3$, so that (\ref%
{vw29}) is fulfilled.

\subsection{Description of polar systems in the $\mathbf{E}$-ensemble} \label{perm-dip}

In this subsection we will generalize the statistical mechanical description of polarizable liquids in an electric field developed above to polar liquid systems.
As induced dipole moments are in many cases much smaller than the permanent ones, we neglect the polarization effects.
However, a generalization of the formalism taking into account both the effects of permanent and induced dipole moments is rather straightforward.
In particular, the latter is expected to be of interest for the computation of the dielectric constant for polar liquids.

The interaction energy is now given by $H_{el,d}=H_{dd}-\sum_i p_iE_0(r_i)$ with
\begin{equation}
H_{dd}=\sum_{i,j}p_i^\alpha T^{\alpha\beta}(r_i-r_j)p_j^\beta \label{dd1}
\end{equation}
being the energy of dipole-dipole interactions.
The configurational integral in the $\mathbf{E}$-ensemble is defined similarly to Eq.~(\ref{td5}), where $H_{pol}$ in the expression of $Q_N(E_0)$ in Eq.~(\ref{vw2}) is replaced by $H_{el,d}$. The functional integration over $E_0(r)$ results instead of (\ref{td5c}) in the following exact expression
\begin{eqnarray}
&&\tilde{Q}_{N}(\mathbf{E})= \int d\Gamma \exp \left( -\beta H_{0}\right)
\exp \left[ \frac{1}{2}\ln (\frac{8\pi ^{2}}{\beta }\det A_0^{-1})\right]
\notag \\
&&\times  \int d^{3}p_{1}\cdots \int d^{3}p_{N}\notag \\
&&\times\exp \left[ -\beta H_{dd}+\frac{\beta }{8\pi }\int d^{3}r\left(\mathbf{E}(r)+4\pi \mathbf{P}(r) \right)^2\right],  \label{dd2}
\end{eqnarray}
where $A_0(r-r')=\delta(r-r')$, and $\mathbf{P}(r)=\sum_i\mathbf{p}_i\delta(r-r_i)$ is the density of the dipole moment.

To enable a quantitative conclusion on the effect of dipole-dipole interactions we restrict ourselves here to the calculation of $\tilde{Q}_{N}(\mathbf{E})$ to the first order in $H_{dd}$. One obtains immediately from (\ref{dd2})
\begin{eqnarray}
&&\tilde{Q}_{N}(\mathbf{E})=\tilde{Q}_p(E)\int d^{3}r_{1}\cdots \int d^{3}r_{N}e^{-H_0}\left(1-\right. \notag  \\
&& \left. \beta\sum_{i,j}<p_i^\alpha>T^{\alpha\beta}(r_i-r_j)<p_j^\beta>+\cdots \right),
\label{dd3}
\end{eqnarray}
where $\tilde{Q}_p(E)=(4 \pi  \sinh (\chi )/\chi )^N$ with $\chi=\beta pE$ is the partition function of the non interacting dipole moments in an electric field, and $<p_i>=p \left(-1/\chi+\coth (\chi )\right)$ is the absolute value of the mean dipole moment. The calculation of (\ref{dd3}) occurs similar to that in (\ref{vw7}). The contribution to the free energy can be obtained from Eq.~(\ref{vw8}) by the replacement $\alpha E_0 \rightarrow <p_i>\equiv<p(E)>$, and yields
\begin{equation}
\tilde{\mathcal{F}}(E)=F_0+\tilde{\mathcal{F}}_p(E)+N\frac{4\pi}{3}<p(E)>^2\frac{N}{V}+\cdots,
\label{dd4}
\end{equation}
where $\tilde{\mathcal{F}}_p(E)=-k_BT\ln\tilde{Q}_p(E)$.
The contribution of permanent electric dipoles to the binary interaction energy in the leading order is given by
\begin{equation}
W_{2,dd}=\frac{1}{2}\sum_{i,j}<p_i^\alpha>T^{\alpha\beta}(r_i-r_j)<p_j^\beta>.\label{dd5}
\end{equation}
The contribution of $W_{2,dd}$ to the Ornstein-Zernike formula can be obtained from (\ref{dd5} in a straightforward way, and is similar to Eq.~(\ref{vw28}). The last term in Eq.~(\ref{dd4}) is similar to that for the free energy in the $\mathbf{E}_0$-ensemble given by Eq.~(\ref{vw8}), and results, as we will see in the next section, in a downward shift of the critical temperature.

\section{Shift of the critical temperature in an electric field} \label{T_c-shift}

Using the above results we will consider the shift of the critical point of a simple liquid in an electric field. The pressure $p=-\left( \partial \mathcal{F}/\partial V\right) _{T,\mathbf{E}_{0}}$ (not to be confused with the absolute value of the permanent dipole moment) is
obtained from Eq.~(\ref{vw8}) as
\begin{equation}
p=p_{0}+\frac{4\pi }{3}\frac{N^{2}}{V^{2}}\alpha _{s}^{2}\mathbf{E}_{0}^{2},
\label{vw15}
\end{equation}
where the pressure $p_{0}$ in the reference state can be identified with
that of the van der Waals equation. The last term in (\ref{vw15})
corresponds to the dipole-dipole interactions of induced dipoles, which are repulsive according to (\ref{vw15}). The interaction of the induced dipoles to the lowest order in $\alpha_s$ ($=\alpha _{s}\mathbf{E} _{0}$) with an homogeneous field does not depend on volume, and thus does not contribute to the pressure. For a dielectric material in a parallel plate capacitor this case corresponds to the charged capacitor in a open circuit. For computation of the pressure in the $\mathbf{E}$-ensemble we use the relation $\tilde{p}=-\left(
\partial \tilde{\mathcal{F}}/\partial V\right) _{T,\mathbf{E}}$ with $\tilde{\mathcal{F}}=\tilde{F}+\int \mathbf{E}^2(\mathbf{r})d^3r/8\pi$ (the latter is the energy of the external field), and $\tilde{F} \simeq -k_BT\ln\tilde{Q}_N(\mathbf{E})$ with $\tilde{Q}_N(\mathbf{E})$ defined by Eq.~(\ref{td5}),
and obtain to order $\alpha_s^2$
\begin{equation}
\tilde{p}=p_{0}-\frac{2\pi }{3}\frac{N^{2}}{V^{2}}\alpha _{s}^{2}\mathbf{E}
^{2}.  \label{vw15a}
\end{equation}

It follows from Eqs.~(\ref{vw15},\ref{vw15a}) that the shift of the critical point in an electric field depends on the ensemble, i.e. if the electric circuit is open or closed in the example with parallel plate capacitor. The
contribution of the electric field to $p$ in Eq.~(\ref{vw15}), which is due
to the dipole-dipole interactions of induced dipoles, is positiv, and
diminishes the constant $a$ of the van der Waals equation, and thus results in a decrease of the critical temperature $T_{c}$. It follows from Eq.~(\ref{vw15a}) that the contribution of the electric field to the pressure at constant $\mathbf{E}$ increases the value of the constant $a$, so that the shift of the critical temperature is upward in the $\mathbf{E}$-ensemble. Note that $T_{c}$ for the van der Waals equation is given by the expression $k_{B}T_{c}=8a/27b$. With the shift of $a$ derived from Eq.~(\ref{vw15a}), $\delta a_{E}=(2\pi /3)\alpha _{s}^{2}\mathbf{E}^{2}$, we obtain the shift of
the critical temperature as $ \delta T_{c}=16\pi \rho _{c}\alpha _{s}^{2}%
\mathbf{E}^{2}/27k_{B}$. The latter coincides with the expression given by Landau and Lifshitz \cite{landau-lifshitz8} $ \delta T=\frac{\rho E^2}{8\pi}\left(\frac{\partial^2 \varepsilon}{\partial
\rho^2}\right)_T/ \frac{\partial ^2p}{\partial\rho\partial T}$
if evaluated with the Clausius-Mossotti formula and the van der Waals
equation. The upward shift in the $\mathbf{E}$-ensemble can be understood as follows. The macroscopic electric field is given by averaging Eq.~(\ref{E-m}). The uniformity of $\mathbf{E}$ is ensured by fluctuations of the external field, which compensate the fluctuations of the electric field of induced dipoles
\begin{equation*}
\delta E_{0}\simeq \delta \left( \sum\limits_{j}T(r-r_{j})\alpha
_{j}E_{0}\right),
\end{equation*}
which are due to the thermal motion of atoms. Thus, the interactions of the induced dipoles with the external field, $\simeq -p_{i}\delta E_{0}$ with $
p_{i}=\alpha _{i}E_{0}$ and $\delta E_{0}$ given above, are negative, and are of the same order of magnitude as the dipole-dipole interactions. The first term on the right-hand side of the effective binary energy in Eq.~(\ref{W2,el}) originates from this term. It follows from Eq.~(\ref{W2,el}) or Eq.~(\ref{vw35}) that the total sign of the effective binary interactions is determined by the first term. In the $\mathbf{E} $-ensemble the interactions of the induced dipoles with the electric field enforce spontaneous inhomogeneities of the density, leading to an increase of the electric energy. This is similar to the behavior in an inhomogeneous electric field, which favors demixing \cite{tsori-nature04}.

The shift of the critical temperature temperature for polar liquids can be obtained from Eq.~(\ref{dd4}) similar to the derivation of the $T_c$ shift in the $\mathbf{E}_0$-ensemble at the beginning of the Section \ref{T_c-shift}. The contribution to the pressure is obtained from Eq.~(\ref{vw15}) by the replacement $\alpha E_0 \rightarrow<p(E)>$. Thus, the permanent dipole moments result in a downward shift of the critical temperature. As the average dipole moment in (\ref{dd4}) equals for moderate field $Ep^2 \beta/3$ the downward shift of the critical temperature is proportional to the square of the field strength.

The results of this paper show that the sign of the shift of the critical temperature is determined by the permanent dipole moments of the atoms or molecules. Under the condition $p/\alpha E> 1$ the shift of $T_c$ is downward for polar systems and upward for unpolar, but polarizable systems. This conclusion is in accordance with the available experimental results in \cite{debye-kleboth65}, \cite{wirtz-fuller-prl93}, \cite{hegseth-prl04}, \cite{park03}, \cite{schoberth}. The molecule sulfur hexafluoride $(SF_6)$ does not possess a permanent electric dipole moment, so that according to our work an upward shift is expected. Such an upward shift for $SF_6$ was established in \cite{hegseth-prl04}. A downward shift for nitrobenzene ($C_6H_5NO_2$), which possesses a permanent dipole moment of $4.23$ D, was measured in \cite{orzechowski99}, and in the earlier work \cite{debye-kleboth65}, and is also in accordance with our prediction. A downward shift was also measured in \cite{wirtz-fuller-prl93} for polystyrene/cyclohexane system. Styrene monomers have a dipole moment of $0.13$ D, while the cyclohexane molecules are unpolar, so that the negative $T_c$ shift in this system is also in accordance with our prediction. In the recent paper \cite{schoberth} a downward shift of the disorder to order transition in polystyrene-block-polyisoprene diblock copolymers in concentrated solutions was measured. For fields as high as $8.5$ kV/mm a decrease in $T_c$ by more than $1.5$ K was observed.
The dipole moments of styrene and isoprene monomers are equal to $0.13$ D and $0.38$ D, respectively, which explains the downward shift.

The estimate of the electric energies of the isoprene monomers due to the polarization effect and the permanent dipole moment $p$ in \cite{boeker-natmat07} yields for the ratio $p/\alpha E\gg 1$, and thus legitimate the neglection of induced electric dipole moments in Section \ref{perm-dip}.

\section{Conclusions}\label{conclus}

To conclude, the statistical mechanical description of liquid systems in an electric field in the ensemble at constant potential, which we establish first for polarizable liquids, provides the basis for systematic calculations of thermodynamic quantities, dielectric constants, nonlinear effects in $\mathbf{E}$ beyond $\mathbf{E}^2$ for broad class of liquid systems like mixtures, polymers, etc.
The generalization of this formalism to liquid systems with permanent dipole moments enables us to analyze the shift of critical temperature in an electric field in polarizable and polar liquids. Our prediction that the shift is upward for polarizable systems, and downward for polar liquids is in accordance with available experimental results.

\begin{acknowledgments}
Financial support from the Deutsche Forschungsgemeinschaft, SFB 418 is
gratefully acknowledged. We would like to thank the Referees for bringing our attention to the paper \cite{schoberth}, and A. B\"{o}ker for sending us the preprint \cite{schoberth} prior to publication.
\end{acknowledgments}

\end{document}